\documentclass{jfm}

\usepackage{amssymb,amsfonts,amsmath}
\usepackage{graphicx}
\usepackage{natbib}
\def\pct{\%{ }}

\def\ie{\emph{i.e.}}
\def\eg{\emph{e.g.}}
\def\cf{\emph{cf.}}

\title{Coherent Lagrangian vortices: The black holes of turbulence}

\author[G. Haller and F. J. Beron-Vera]{G.\ Haller$^1$\thanks{Email
address for correspondence: georgehaller@ethz.ch}\ns and F.\ J.\
Beron-Vera$^2$}

\affiliation{$^1$Institute for Mechanical Systems, ETH Zurich,
Zurich, Switzerland\\[\affilskip] $^2$ Rosenstiel School of Marine
and Atmospheric Science, University of Miami, Miami FL, USA}

\pubyear{2013}
\volume{X}
\pagerange{X--X}
\date{13 May 2013; revised 18 July 2013; accepted 23
July 2013.}
\begin{document}

\maketitle

\begin{abstract}
  We introduce a simple variational principle for coherent material
  vortices in two-dimensio- nal turbulence. Vortex boundaries are
  sought as closed stationary curves of the averaged Lagrangian
  strain. Solutions to this problem turn out to be mathematically
  equivalent to photon spheres around black holes in cosmology. The
  fluidic photon spheres satisfy explicit differential equations
  whose outermost limit cycles are optimal Lagrangian vortex
  boundaries.  As an application, we uncover super-coherent material
  eddies in the South Atlantic, which yield specific Lagrangian
  transport estimates for Agulhas rings.
\end{abstract}

\section{Introduction}

Vortices in turbulence are often envisaged as rotating bodies of
fluid, traveling as coherent islands in an otherwise incoherent
ambient flow \citep{Provenzale-99}. This Lagrangian view is appealingly
simple, yet challenging to apply in actual vortex detection. The
main difficulty is to classify fluid particle paths systematically
as coherent or incoherent.

As a result, coherent features of turbulence are mostly described
through instantaneous Eulerian quantities, such as velocity, pressure
and their derivatives \citep[\cf][for reviews]{Jeong-Hussain-95,
Haller-05}.  In this approach, vortices are typically defined as
regions of closed contours of an Eulerian scalar field, often with
some added thresholding requirement. Such criteria can be highly
effective in framing short-term features of the flow, even if the
extracted coherent structures tend to be frame-dependent.

In unsteady flow, however, Eulerian vortex boundaries are not
material barriers. A fluid mass initialized in an Eulerian vortex
will generally lose coherence, showing stretching and filamentation
along the vortex path. The end-result for the fluid mass is widespread
dispersion with little or no directionality \citep{Beron-etal-13}.
Yet identifying coherent material vortices is becoming increasingly
important in a number of areas. For instance, mesoscale oceanic
eddies are broadly thought to carry water without substantial leakage
or deformation.  If their boundaries indeed resist filamentation,
such eddies can create moving oases for the marine food chain
\citep{Denman-Gargett-83}, or even impact climate change through
their long-range transport of salinity and temperature
\citep{Beal-etal-11}.

Lagrangian diagnostic tools show that some Eulerian vortices can
carry fluid, but their leakage tends to be substantial
\citep{Provenzale-99, Froyland-etal-12}.  A rigorous approach to
finding non-leaking material boundaries has recently emerged, but
focuses only on relatively rare boundaries that maximize Lagrangian
shear \citep{Beron-etal-13}.

\begin{figure}
  \centering%
  \includegraphics[width=\textwidth]{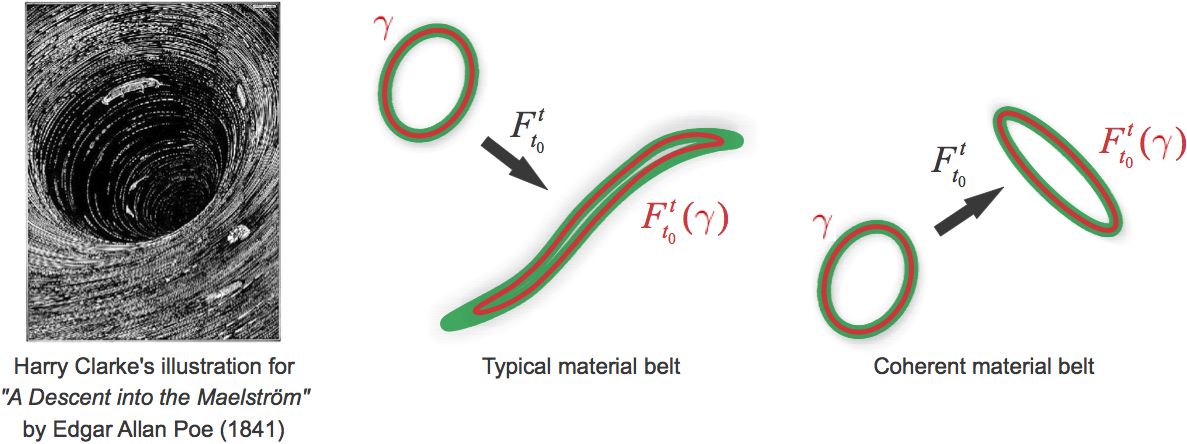}%
  \caption{Edgar Allan Poe's maelstrom and material belts in
  turbulence. A closed material curve $\gamma$ (red) at time $t_0$
  is advected by the flow into its later position $F^t_{t_0}(\gamma)$
  at time $t$. The advected curve remains coherent if an initially
  uniform material belt (green) around it shows no leading-order
  variations in stretching after advection. \label{fig:deffigure}}
\end{figure}

Parallel to these developments, coherent vortices are sometimes
described as whirlpools, or maelstroms, in popular fiction. An early
example can be found in Edgar Allan Poe's short story entitled
\emph{A Descent into a Maelstr\"om}:
\begin{quotation}
  ``The edge of the whirl was represented by a broad belt of gleaming
  spray; but no particle of this slipped into the mouth of the
  terrific funnel\dots''
\end{quotation}
This literary account depicts a belt-like vortex boundary
that keeps particles from entering its interior (Figure
\ref{fig:deffigure}).  Altogether, Poe's view on vortices is
Lagrangian, and resonates with our intuition for black holes in
cosmology.

As we show below, this view turns out to have some merit. When
appropriately modeled, Poe's coherent belt becomes mathematically
equivalent to a photon sphere, \ie, a surface on which light encircles
a black hole without entering it. This analogy yields computational
advantages, which we exploit in locating material eddy boundaries
in the South Atlantic Ocean. Using satellite altimetry-based
velocities from this region, we uncover super-coherent Lagrangian
vortices, and derive estimates for coherent material transport
induced by the Agulhas leakage.

\section{Coherent material belts}

We start with a two-dimensional velocity field of the form $v(x,t),$
with $x$ labeling the location within an open region $U$ of interest
and with $t$ referring to time. Fluid trajectories generated by
$v(x,t)$ obey the differential equation
\begin{equation}
  \dot{x} = v(x,t),
  \label{eq:uxt}
\end{equation}
whose solutions are denoted $x(t;t_{0},x_{0}$), with $x_{0}$ referring
to the initial position of the trajectory at time $t_{0}$. The
evolution of fluid elements is described by the flow map
\begin{equation}
  F_{t_{0}}^{t}(x_{0}) := x(t;t_{0},x_{0}),
\end{equation}
which takes any initial position $x_{0}$ to a later position at
time $t$. Lagrangian strain in the flow is often characterized by
the right Cauchy--Green strain tensor field $C_{t_{0}}^{t}(x_{0})
= \nabla F_{t_{0}}^{t}(x_{0})^\top\nabla F_{t_{0}}^{t}(x_{0})$,
whose eigenvalues $\lambda_{i}(x_{0})$ and eigenvectors $\xi_{i}(x_{0})$
satisfy:
\[
  C_{t_{0}}^{t}\xi_{i} = \lambda_{i}\xi_{i},\quad
  \left|\xi_{i}\right|=1,\quad i=1,2;\quad
  0<\lambda_{1}\leq\lambda_{2},\quad
  \xi_{1}\perp\xi_{2}.
\]

Assume that $\varepsilon>0$ is a minimal threshold above which we
can physically observe differences in material strain over the time
interval $[t_{0},t].$ By smooth dependence on initial fluid positions
\citep{Arnold-73}, we will observe an $O(\varepsilon)$ variability
in strain within an $O(\varepsilon)$ belt around a generic closed
material curve $\gamma$.  However, we seek exceptional $\gamma$
curves around which $O(\varepsilon)$-thick coherent belts show no
observable variability in their average straining (Figure
\ref{fig:deffigure}).

Such coherence in the strain field precludes nearby fluid elements
from breaking away from $\gamma$. Again, by smooth dependence of
finite-time strain on distance, the actual strain variability in
such coherent, $O(\varepsilon)$-thick belts cannot be more than
$O(\varepsilon^{2})$.

To express this coherence principle mathematically, we select a
parametrization $r(s)$ with $s \in [0,\sigma]$ for the closed curve
$\gamma$ at time $t_{0}$, and denote the length of a tangent vector
$r^{\prime}(s)$ by $l_{t_{0}}(s)$ . We also let $l_{t}(s)$ denote
the length of the corresponding tangent vector
$\frac{\mathrm{d}}{\mathrm{d}s}F_{t_{0}}^{t}(r(s))$ along the
advected curve $F_{t_{0}}^{t}(\gamma)$. These two tangent lengths
can be calculated as
\begin{equation}
  l_{t_{0}}(s) = \sqrt{\langle r^{\prime}(s),r^{\prime}(s)\rangle},\quad
  l_{t}(s) = \sqrt{\langle
  r^{\prime}(s),C_{t_{0}}^{t}(r(s))r^{\prime}(s)\rangle},
\end{equation}
where $\langle\cdot,\cdot\rangle$ denotes the Euclidean inner product
\citep{Truesdell-Noll-04}. The averaged tangential strain along $\gamma$ is then given by
\[
  Q(\gamma) = \frac{1}{\sigma}\int_{0}^{\sigma}
  \frac{l_{t}(s)}{l_{t_0}(s)}\,\mathrm{d}s.
\]
As argued above, if an observable coherent material belt exists
around $\gamma$, then on $\varepsilon$-close material loops we must
have $Q(\gamma + \varepsilon h) = Q(\gamma) + O(\varepsilon^{2})$,
where $\varepsilon h(s)$ denotes small, periodic perturbations to
$\gamma$.   This is only possible if the first variation of $Q$
vanishes on $\gamma$:
\begin{equation}
  \delta Q(\gamma) = 0.\label{eq:zerovari}
\end{equation}

The calculus of variations applied to \eqref{eq:zerovari} leads to
complicated differential equations that do not immediately provide
insight (Appendix A). However, as we show in Appendix B, any $\gamma$
satisfying \eqref{eq:zerovari} also satisfies
\begin{equation}
  \delta\mathcal{E}_{\lambda}(\gamma) = 0,\quad
  \mathcal{E}_{\lambda}(\gamma) = \oint_{\gamma}\langle
  r^{\prime}(s),E_{\lambda}(r(s))r^{\prime}(s)\rangle\,\mathrm{d}s,
  \label{eq:qdef-1}
\end{equation}
representing a closed stationary curve for the strain energy
functional $\mathcal{E}_{\lambda}(\gamma)$ for some choice of the
parameter $\lambda > 0$. Here the tensor family
\begin{equation}
  E_{\lambda}(x_{0}) = \frac{1}{2}\left[C_{t_{0}}^{t}(x_{0}) -
  \lambda^{2}I\right] 
  \label{eq:Elambda}
\end{equation}
denotes a generalization of the classic Green--Lagrange strain
tensor $E_{1}(x_{0}) = \frac{1}{2}[C_{t_{0}}^{t}(x_{0})-I]$
\citep{Truesdell-Noll-04}.

\section{Lagrangian vortex boundaries}

All stationary curves of $\mathcal{E}_{\lambda}$ that coincide with
those of $Q$ are found to have zero energy density (Appendix B),
satisfying the implicit differential equation
\begin{equation}
  \langle r^{\prime}(s),E_{\lambda}(r(s))r^{\prime}(s)\rangle =
  l_{t}^{2}(s)-\lambda^{2}l_{t_{0}}^{2}(s) \equiv 0.  
  \label{eq:defnull}
\end{equation}
These material curves $r(s)$ are therefore uniformly stretched by
the same factor $\lambda$ when advected by the flow from time $t_{0}$
to time $t$. Such $\lambda$\emph{-lines} serve as perfectly coherent
cores for observably coherent material belts around them.

Equation \eqref{eq:defnull} only admits nondegenerate $r(s)$ solutions
on the flow domain $U_{\lambda}$ where the two-dimensional tensor
$E_{\lambda}$ has nonzero eigenvalues of opposite sign. In this
domain the curves $r(s)$ are obtained as closed trajectories of one
of the two explicit differential equations (Appendix C)
\begin{equation}
  r^{\prime}(s) = \eta_{\lambda}^{\pm}(r(s)),\quad% 
  \eta_{\lambda}^{\pm} = \sqrt{\frac{\lambda_{2} - \lambda^{2}}{\lambda_{2}
  - \lambda_{1}}}\,\xi_{1}\pm\sqrt{\frac{\lambda^{2} -
  \lambda_{1}}{\lambda_{2} - \lambda_{1}}}\,\xi_{2}.  
  \label{eq:ode-1}
\end{equation}

Such closed trajectories will occur in families of limit cycles
parametrized by the stretching parameter $\lambda$. One can show
that no two members of any such family can intersect (Appendix E).
Thus a limit cycle of the vector field $\eta_{\lambda}^{\pm}$ will
either grow or shrink under changes in $\lambda$, forming smooth
annular regions of nonintersecting loops. The outermost member of
such a band of coherent material loops will be observed physically
as the Lagrangian vortex boundary. Indeed, immediately outside this
boundary, no coherent material belts may exist in the flow.

\section{Lagrangian vortices and black holes}

The trajectories of \eqref{eq:ode-1}, as stationary curves of the
stretch energy $\mathcal{E}_{\lambda}(\gamma)$, admit an appealing
geometric interpretation. First, note that on the domain $U_{\lambda}$,
the quadratic function
\begin{equation}
  g_{\lambda}(u,u) = \left\langle u,E_{\lambda}u \right\rangle
\end{equation}
defines a Lorentzian metric, with signature $(-,+)$ inherited from
the eigenvalue configuration of $E_{\lambda}$ \citep{Beem-etal-96}.
A Lorentzian metric extends the classic notion of distance to
hyperbolic spaces. As such, it can also return zero or a negative
value for the distance between two points that have a positive
Euclidean distance.

Thus, $(U_{\lambda},g_{\lambda})$ can be viewed as a two-dimensional
Lorentzian manifold \citep{Beem-etal-96}, or in the language of
general relativity, a two-dimensional spacetime. In such a spacetime,
light travels along null-geodesics of the Lorentzian metric, which
are curves on which the Lorentzian metric vanishes identically. In
our fluidic spacetime $(U_{\lambda},g_{\lambda})$, null-geodesics
satisfy $g_{\lambda}(r^{\prime}(s),r^{\prime}(s))\equiv0,$ and hence
coincide with the $\lambda$-lines described in \eqref{eq:defnull}.

\begin{figure}
  \centering%
  \includegraphics[width=.75\textwidth]{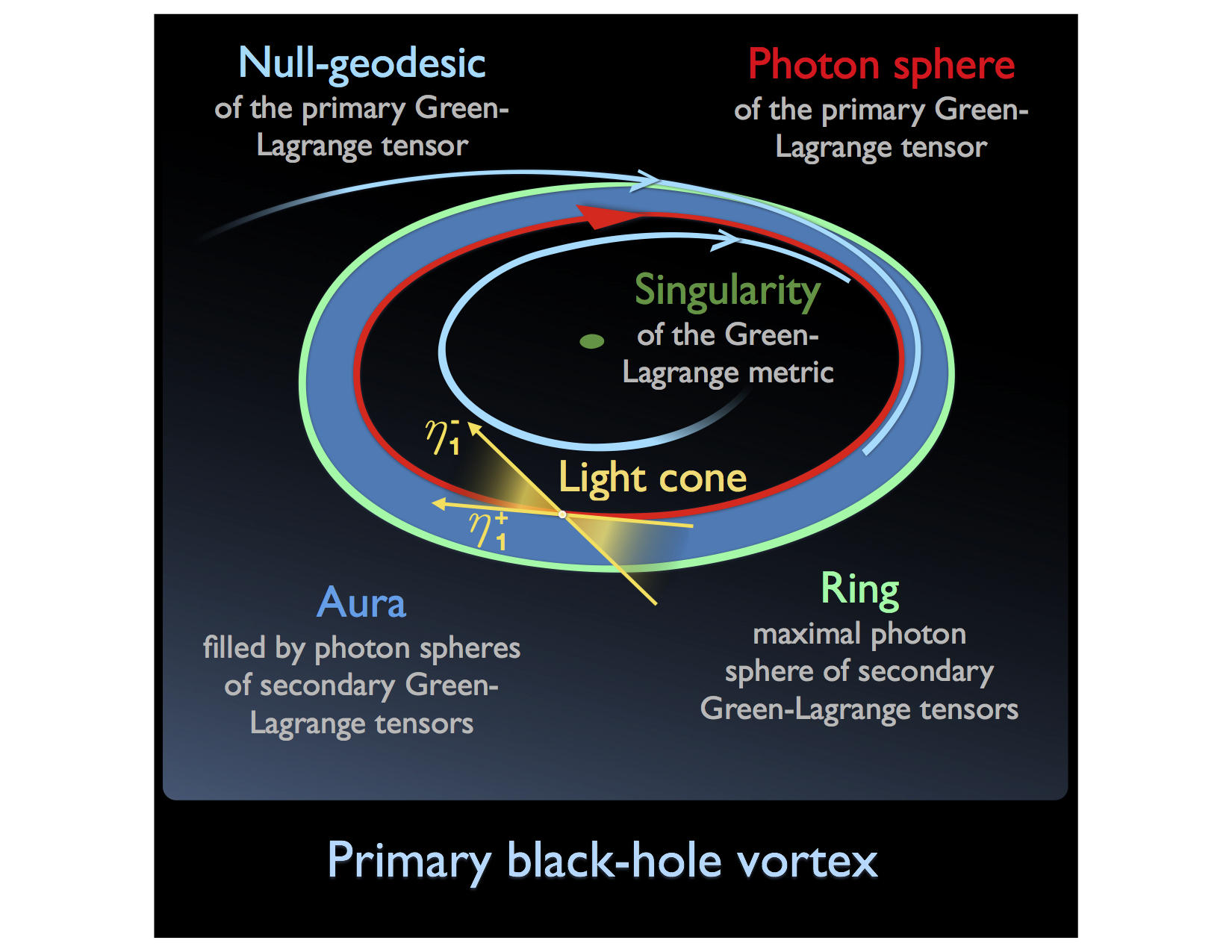}%
  \caption{Mathematical equivalence between coherent Lagrangian
  vortices and black holes.\label{fig:analogy}}
\end{figure}

The closed $\lambda$-lines we have been seeking are therefore closed
null-geodesics of $g_{\lambda}$. In cosmology, such null-geodesics
occur around black holes, representing loops that trap light forever.
Often called \emph{photon spheres} \citep{Beem-etal-96}, these
closed curves are tangent to the \emph{light cones}, which in our
case are defined by the two vectors $\eta_{\lambda}^{\pm}(x_{0})$
(Figure \ref{fig:analogy}).

Due to its special nonstretching nature, we refer to a photon sphere
of $E_{1}$ as a \emph{primary photon sphere}. Photon spheres of
$E_{\lambda}$ with $\lambda \neq 1$ will be referred to as
\emph{secondary photon spheres}. A primary photon sphere in our
context resists the universally observed material stretching in
turbulence: it reassumes its initial arclength at time $t$. This
conservation of arclength, along with the conservation of the
enclosed area in the incompressible case, creates extraordinary
coherence.

Even if a primary photon sphere does not exist in a region, secondary
photon spheres with either $\lambda>1$ or $\lambda<1$ may well be
present. The largest of these photon spheres marks the end of the
family of coherent belts. It will signal either weak stretching
($\lambda >1$) or weak contraction ($\lambda <1$) for the vortex
it bounds. Weakly compressing Lagrangian vortices are evolving
towards even fuller coherence, preserving their enclosed area while
smoothing out mild bumps in their boundaries. By contrast, weakly
stretching Lagrangian vortices can be viewed as domains of slowly
eroding coherence, where a perfect, nonstretching core can no longer
be found.

Motivated by these observations, we introduce the following
terminology.  A \emph{black-hole} \emph{vortex} is a region surrounded
by an outermost (i.e., locally the largest) photon sphere (or
\emph{ring}) of the tensor family $E_{\lambda}$.  A \emph{primary
black-hole vortex} is one that contains at least one primary photon
sphere, such as the one shown in Figure \ref{fig:analogy}. A
\emph{secondary black-hole vortex} is one that only contains secondary
photon spheres. Such a vortex is either \emph{strengthening}
($\lambda<1)$ or \emph{weakening} ($\lambda>1)$.

\section{Detection of Lagrangian vortices}

The above analogy with black holes has practical implications for
locating Lagrangian vortices in a complex flow. To see this, note
that the strain eigenvectors $\xi_{1}$ and $\xi_{2}$ in \eqref{eq:qdef-1}
become ill-defined at points where $\lambda_{1}(x_{0}) =
\lambda_{2}(x_{0})=1$.  Null-geodesics, such as those forming photon
spheres, cannot be extended to such points, because the Green--Lagrange
metric $g_{1}$ is singular there [$E_{1}(x_{0}) = 0$]. These
degenerate points are analogs of Penrose--Hawking singularities in
cosmology \citep{Hawking-Penrose-96}.  Intuitively, one expects
that any coherent Lagrangian vortex in the fluid must contain such
a singularity in its interior, just as all black holes are expected
to contain Penrose--Hawking singularities. This expectation turns
out to be correct, leading to the following result (Appendix D):
at time $t_{0}$, the interior of a Lagrangian vortex must contain
a point $x_{0}^{*}$ satisfying $E_{1}(x_{0}^{*})=0$.  Specifically,
only regions containing such points may fully incorporate Lagrangian
vortices.

Based on these findings, admissible regions for coherent Lagrangian
vortices are those with at least one metric singularity [$E_{1}(x_{0}^{*})
= 0$].  In practice, metric singularities tend to cluster, and we
only consider clusters surrounded by an annular region with no
singularities. In these admissible regions, Lagrangian vortex
boundaries are sought as outermost Green--Lagrange photon spheres,
\ie, outermost limit cycles of the differential equations
\eqref{eq:ode-1}. This process involves conceptually simple numerical
steps, but its implementation requires care for noisy observational
data (Appendix F).

\section{Lagrangian vortices in the Agulhas leakage}

We apply our results to the Agulhas leakage, a byproduct of the
Agulhas current of the southwest Indian Ocean. At the end of its
southward flow, this boundary current turns back on itself, creating
a loop that occasionally pinches off and releases eddies (Agulhas
rings) into the South Atlantic.

Lagrangian transport by coherent Agulhas rings has implications for
global circulation and climate change \citep{Beal-etal-11}. As shown
by \citet{Beron-etal-13}, eddy tracking from instantaneous altimetric
Sea-Surface-Height (SSH) measurements \citep{Chelton-etal-11a} would
substantially overestimate the coherent part of this transport.
Here we reconsider the region studied by \citet{Beron-etal-13} and
show how the coherent Lagrangian vortex principle developed here
uncovers previously unknown material eddies, sharpening available
coherent transport estimates. The South Atlantic ocean region in
question is bounded by longitudes [14$^{\circ}$W, 9$^{\circ}$E] and
latitudes [39$^{\circ}$S, 21$^{\circ}$S].  Using satellite altimetry
data, we seek coherent Lagrangian vortices (black-hole eddies, for
short) over a 90-day time period, ranging from $t_{0}$ = 24 November
2006 to $t=$ 22 February 2007. Appendix G provides more information
on the data and the numerical methods employed.

\begin{figure}
  \centering% 
  \includegraphics[width=\textwidth]{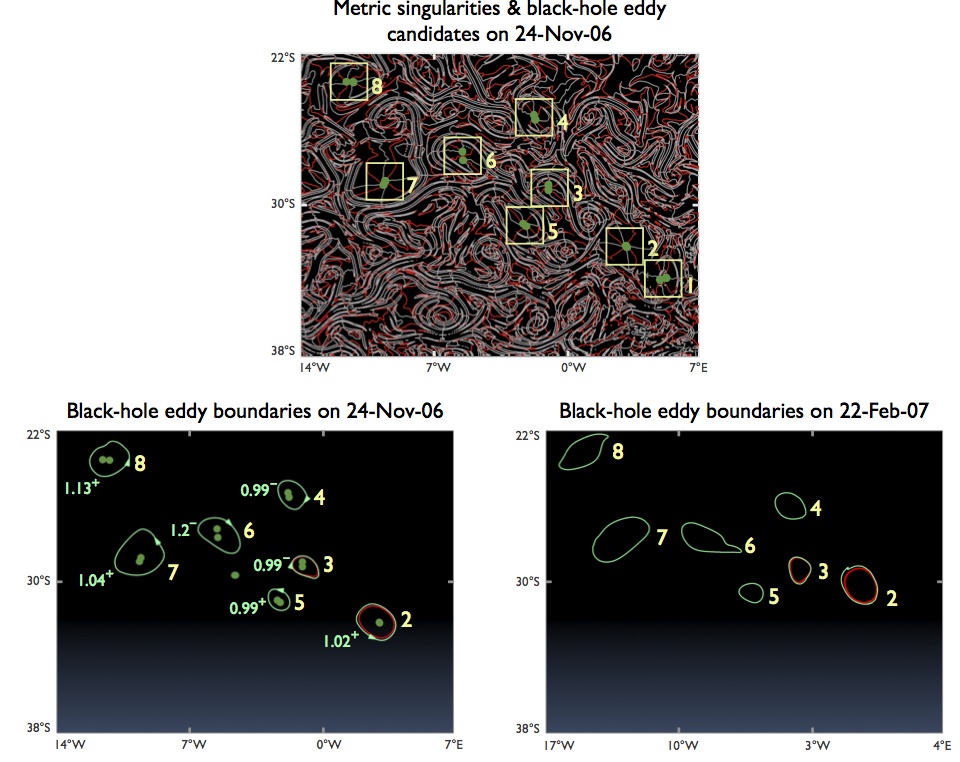}%
  \caption{(top panel) Green--Lagrange singularities [intersections
  of red and white curves; \cf Appendix F, step (\emph{c})] and
  admissible Lagrangian vortex regions (yellow). The dark green
  singularity clusters are surrounded by singularity-free belts of
  at least 50-km in diameter.  (bottom-left panel) Black-hole eddy
  boundaries (light green, with $\lambda$ values in light green)
  and primary photon spheres (red). The $\pm$ signs shown are those
  of $\eta_{\lambda}^{\pm}$. (bottom-right panel) Advected Lagrangian
  vortex boundaries $3$ months later.  \label{fig:BH-candidates}}
\end{figure}

The computational steps reviewed in the Appendix F yield eight
admissible regions for Lagrangian vortices (Figure \ref{fig:BH-candidates},
top panel).  Out of these eight regions, only seven turn out to
contain black-hole eddies, bounded by outermost Green--Lagrange
photon spheres (Figure \ref{fig:BH-candidates}, bottom-left panel).
Two of these are primary black-hole eddies, containing primary
photon spheres. The latter curves were previously identified by
\citet{Beron-etal-13} as locally most shearing lines from other
methods.

Out of the two primary black-hole eddies, eddy 2 has a slightly
weakening external boundary with $\lambda=1.02$, while eddy 3 has
a slightly strengthening boundary with $\lambda=0.99$. The boundaries
of the secondary black-hole eddies 4 and 5 are also slightly gaining
coherence ($\lambda=0.99$), whereas the secondary eddies 6, 7 and
8 are losing coherence ($\lambda=1.2$, 1.04 and 1.13, respectively).
Accordingly, the boundary of eddy 6 is expected to exhibit the
largest degree of stretching. This stretching is still uniform,
leaving room for a coherent material belt for the 90-day study
period. These conclusions are confirmed by the bottom-right panel
of Figure \ref{fig:BH-candidates}, showing the seven advected
black-hole eddy boundaries three months later.

We verified the theoretically predicted $\lambda$ stretching values
for black-hole eddy boundaries by their direct numerical advection.
In this test, we found the $\lambda$ values obtained from advection
to be accurate up to two decimal digits, as  listed in the left-bottom
panel of Figure \ref{fig:BH-candidates}.  Since these eddy boundaries
are constructed as limit cycles, they are structurally stable, i.e.,
smoothly persist under small errors and uncertainties in the
observational data set.

The detailed knowledge of black-hole eddies yields specific estimates
for their Lagrangian transport rates. As in \citet{Goni-etal-97},
we assume that the eddies are bounded from below by the isotherm
of 10$^{\circ}$C, located roughly at 400 m in the area of study.
Under this assumption, we obtain the eddy transport rates shown in
Table \ref{tab:sv}. On average, a black-hole eddy moves water
northwestward at a rate of about 1.3 Sv (1 Sv $=10^{6}$ m$^{3}$s$^{-1}$).
This represents a 30\pct upward refinement to the 1-Sv estimate of
\citet{Goni-etal-97} for average transport by SSH eddies in the
Agulhas leakage. More significant is the difference between the
total eddy transport calculated from black-hole eddies and the total
eddy transport calculated from SSH eddies using the estimate of
\citet{Goni-etal-97}. As seen from Table \ref{tab:sv}, SSH snapshots
significantly overestimate coherent eddy transport.

\begingroup
\renewcommand*{\arraystretch}{1.35}
\begin{table}
  \centering   
  \begin{tabular}{cccccccccc}%
    \hline%
     \multicolumn{7}{c}{BH eddy} &  &  &  \\ \cline{1-7}%
     2 & 3 & 4 & 5 & 6 & 7 & 8 & BH average & BH total & SSH total \\%
     1.4 & 1.8 & 0.8 & 1.0 & 1.1 & 1.0 & 2.0 & 1.3 & 9.1 & 18 \\%
    \hline%
  \end{tabular}
  \caption{Transport rates in Sv for black-hole (BH)
  eddies and SSH eddies on 24 November 2006. }
  \label{tab:sv}
\end{table}
\endgroup

We note that in situ measurements analyzed by \citet{vanAken-etal-03}
reveal some mesoscale eddies reaching depths as large as 2000 m or
more, as opposed to the generic 500-m depth assumed by \citet{Goni-etal-97}
and in the present work. The difference in the previous Eulerian
and the present Lagrangian transport estimates increases linearly
with the assumed depth for the eddy.

\begin{figure}
  \centering%
  \includegraphics[width=\textwidth]{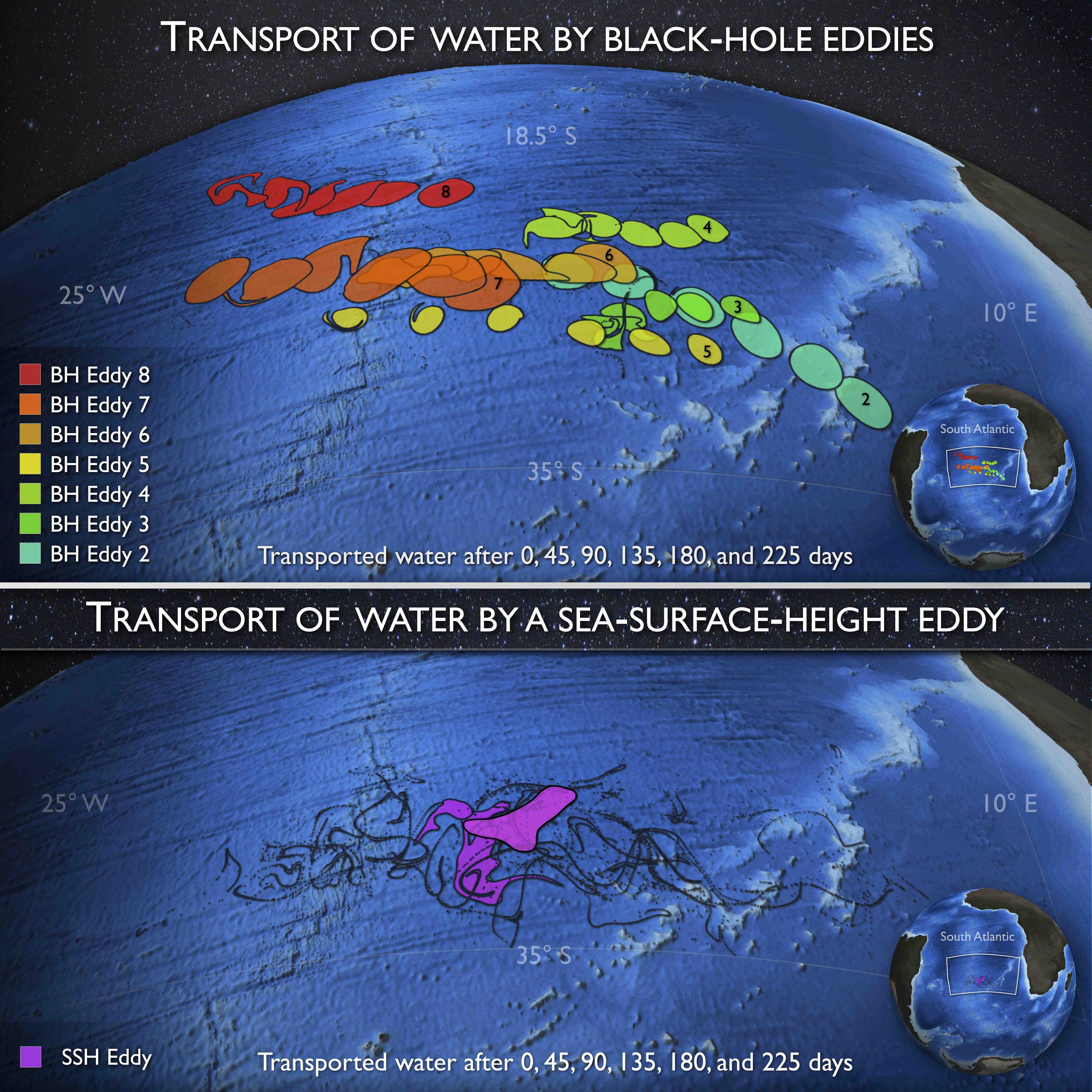}%
  \caption{(top panel) Evolution of black-hole eddies (extracted
  from 3 months of data) in the South Atlantic over a period of 225
  days. The eddies move from east to northwest (form right to
  left). (bottom panel) Material evolution of a nonlinear SSH eddy
  over the same 225 days.\label{fig:BH-tracks}}
\end{figure}

Finally, the top panel of Figure \ref{fig:BH-tracks} shows a longer
computation of the paths of the seven black-hole eddies. Most
preserve their coherence way beyond their 3-month extraction period.
Also shown the evolution of fluid starting from a nonlinear SSH
eddy obtained from the analysis of \citet{Chelton-etal-11a} (Figure
\ref{fig:BH-tracks}, bottom panel). This illustrates the rapid loss
of coherence that fluid initialized in an Eulerian vortex will
generically experience in turbulence.

Common Lagrangian diagnostics of material coherence do not fare
well either.  For instance, the Finite-Time Lyapunov Exponent (FTLE)
field displays ridges spiraling into black-hole eddies in the ocean
region studied here.  This incorrectly signals a lack of coherent
material boundaries for Agulhas rings, as noted by \citet{Beron-etal-08b}.

\section{Conclusions}

We have given a variational description of coherent Lagrangian
vortex boundaries as closed belts of fluid showing no leading-order
variability in their averaged material straining. Such material
belts are filled with closed null-geodesics of a generalized
Green--Lagrange strain metric, which can be computed as limit cycles
of a Lagrangian vector field. Outermost members of such limit cycle
families mark observed boundaries of material coherence. Using this
approach, we have found exceptionally coherent material belts in
the South Atlantic, filled with analogs of photon spheres around
black holes.  Our results apply to any two-dimensional velocity
field, providing a frame-independent way to detect coherent Lagrangian
vortices.

In addition to the transport of diffusive water attributes, the
transport of other materials (\eg, garbage or oil contamination)
is also expedited by the coherent Lagrangian vortices identified
here. Related results on elliptic regions in steady inertial particle
motion \citep{Haller-Sapsis-08} suggest that these vortices will
capture and swallow nearby passively floating debris. Thus, beyond
the mathematical equivalence, there are also observational reasons
for viewing coherent Lagrangian eddies as black holes.

Continually available observational data for the ocean are limited
to two spatial dimensions. However, the existence of a coherent
material belt outside down- or upwellings should already be
well-captured by two-dimensional altimetry data. We therefore also
expect black-hole-type vortices to be present in two-dimensional
atmospheric wind data, framing the Lagrangian footprints of hurricanes
on Earth and of the Great Red Spot on Jupiter.

The altimetry data used here are available from AVISO
(http://aviso.oceanobs.com).  GH acknowledges partial support by
NSERC grant 401839-11. FJBV acknowledges support by NSF grant
CMG0825547, NASA grant NNX10AE99G, and by a grant from the British
Petroleum--Gulf of Mexico Research Initiative.

\bibliographystyle{jfm}
%\bibliography{fot}

\end{document}